\documentclass[10pt,twocolumn,twoside]{IEEEtran}

\usepackage[T1]{fontenc}
\usepackage[utf8]{inputenc}

\usepackage[numbers]{natbib}
\usepackage{url}
\usepackage{multirow}
\usepackage{booktabs}
\usepackage{diagbox}
\usepackage{amsmath}
\usepackage{amssymb}
\usepackage{mathtools}

\usepackage{hyperref}
\usepackage{graphicx, grffile}

\usepackage[caption=false,font=footnotesize]{subfig}
\usepackage[table, xcdraw]{xcolor}

\usepackage{longtable}
\usepackage{booktabs}

\usepackage{multicol}

\begin{document}

\title{Project Achoo: A Practical Model and Application for COVID-19 Detection from Recordings of Breath, Voice, and Cough}

    \author{%
        {Alexander~Ponomarchuk}\IEEEauthorrefmark{1},~
        {Ilya~Burenko}\IEEEauthorrefmark{1}\IEEEauthorrefmark{2},~
        {Elian~Malkin}\IEEEauthorrefmark{1},~
        {Ivan~Nazarov}\IEEEauthorrefmark{1},
        {Vladimir~Kokh}\IEEEauthorrefmark{1},~
        {Manvel~Avetisian}\IEEEauthorrefmark{1},~
        {Leonid~Zhukov}\IEEEauthorrefmark{1}%
        
        \IEEEauthorrefmark{1} Sber AI Lab\\  
        \IEEEauthorrefmark{2} Corresponding author: burenko.i.m@sberbank.ru
        }

\date{May 2021}


\maketitle

\begin{abstract}  
%

The COVID-19 pandemic created significant interest and demand for infection detection
and monitoring solutions. In this paper, we propose a machine learning method to quickly
detect COVID-19 using audio recordings made on consumer devices. The approach combines
signal processing and noise removal methods with an ensemble of fine-tuned deep learning
networks and enables COVID detection on coughs. We have also developed and deployed a mobile
application that uses a symptoms checker together with voice, breath, and cough signals
to detect COVID-19 infection. The application showed robust performance on both openly
sourced datasets and the noisy data collected during beta testing by the end users.

\end{abstract}


\IEEEpeerreviewmaketitle

\section{Introduction and Related Work}  
\label{sec:introduction}


\IEEEPARstart{T}{h}e continuing proliferation of smart devices and growth of their computational
power have sparked research interest related to applications of machine learning for audio-based
medical screening of respiratory infections and airborne diseases long before
the COVID-19 pandemic.
The primary motive was to enable affordable data-driven non-invasive health screening methods,
which can be rapidly scaled up in response to public health emergencies and quickly adapted to
particular public health concerns in infection hotspots. This is especially relevant whenever
a high-risk population predominantly lives in remote areas or cases when access to skilled
clinicians or care-takers is limited.

At the early stages of the COVID-19 pandemic, it was important not only to detect patients affected
by the scourge, but also to estimate the severity of the disease by calculating the share of affected
lung tissue, and to distinguish COVID-positive patients from those with other acute respiratory
diseases, e.g. viral or bacterial pneumonia, in order to help doctors prioritize patients and
provide treatment before a PCR result was confirmed.
These reasons led machine learning practitioners to propose diagnostic methods that exploit
visual information, such as chest X-ray images, computer tomography, or ultrasound studies
(see references in \citep{huazhu_fu_covid-19_2021}). Great progress has been made due to
the availability of large datasets, and many clinics around the world have adopted automated
image-based methods to help patients with COVID, \citep{avetisian_corsai_2021}.
However, further spread of the pandemic, coupled with discoveries about COVID's contagiousness
and course, emphasized the necessity of detecting asymptomatic carriers and infected persons
without noticeably affected lung tissue at the early stages of the disease, since, in general,
such patients are not compelled to reduce their social activity and thus could contribute to
the sustained spread of the virus.

The simulation study \citep{larremore_test_2021} has shown that affordable COVID screening, even
if less sensitive than clinical tests, allows to control the spread of the virus more effectively,
by being able to be quickly scaled up and massively deployed.
Furthermore, massive screening could reduce the exposure of first responders and critical
frontline medical workers to virulent respiratory infections, \citep{imran_ai4covid-19_2020}.
%
However, medical imaging studies are impractical for rapid screening purposes and asymptomatic
or unaware carriers are unlikely to undergo such studies or seek out a PCR test. An alternative
could be to use the ``everyday data'' from smart devices, such as samples of voice, breath, and
cough, as an input to an AI non-diagnostic pre-screening tool.
The effectiveness of this approach rests on the hypothesis that at its early stages COVID-19
produces measurable physiological changes, such as sore throat, lung obstruction, or reduced
blood oxygen saturation, \citep{imran_ai4covid-19_2020}.

In general, machine learning and deep learning methods for medical applications require large
datasets with accurate and expertly verified ground truth. The onset of COVID-19 pandemic,
however, has made it challenging to obtain sufficient volumes of high-quality clinically reliable
data, collected under strictly controlled conditions, especially considering time constraints
and logistical limitations.  
In such circumstances, some studies went with the crowdsourcing approach to collecting large
open datasets of coughs for COVID, \citep{orlandic_coughvid_2020,sharma_coswara_2020}, and
\citep{project_fkthecovid_dataset_2021}, (see sec.~\ref{sub:open_datasets}). Crowdsourcing,
however, has severe limitations: lack of adherence to blinding protocols, i.e. the COVID status
is known to the subject prior to participation, poorly controlled selection bias, ambient sound
conditions, or other confounding variables, and, most importantly, unreliable ground-truth
labels, i.e. self-reported COVID status versus a verified PCR test,
\citep{bagad_cough_2020,coppock_covid-19_2021}.

Despite these shortcomings, the available crowdsourced respiratory sound datasets could be
valuable for pre-training deep neural networks for cough detection and COVID identification
tasks, \citep{bagad_cough_2020,laguarta_covid-19_2020}. Thus initialized models can be fine-tuned
on a much smaller dataset of better quality, collected in a controlled setting and having
the infection status verified through repeated properly conducted PCR test, \citep{watson_interpreting_2020}.
This transfer learning approach is adopted in the current study: our solution is pre-trained
on the crowdsourced data with \emph{weak} labels (sec.~\ref{sub:open_datasets}), then
fine-tuned and tested on a privately commissioned higher grade dataset with \emph{strong}
labels (sec.~\ref{sub:private_data}), and additionally validated on the real data, collected
``in the wild'' through a custom mobile application (sec.~\ref{sub:app_data}).

Our key contributions are:
\begin{itemize}
    \item We propose an ensemble of deep convolutional neural networks and gradient
    boosted classifiers that together predict COVID status based on cough recordings
    (sec.~\ref{sec:the_method});
    \item We present the validation results of the proposed pipeline on openly
    accessible datasets, as well as a private dataset collected in COVID wards
    (sec.~\ref{sec:datasets} and sec.~\ref{sec:experiments_and_discussion});
    \item We describe the implementation of the preprocessing and COVID diagnosing
    pipeline in a mobile application developed for rapid COVID screening
    (sec.~\ref{sec:application}).
\end{itemize}

In the following section \ref{sub:related_work} we briefly survey the existing approaches
to cough-based disease detection. Sections \ref{sec:datasets} and \ref{sec:the_method}
describe the used datasets and the proposed approach, respectively, while in section
\ref{sec:experiments_and_discussion} we provide results of our experiments on the private
data along with performance on the public datasets.  Results of our models on crowdsourced
data are presented in section \ref{sec:application}. We share the limitations of our work,
concluding remarks, and outline further research directions in sections \ref{sec:considerations}
-- \ref{sec:conclusion}.

\subsection{Related Work}  
\label{sub:related_work}

Respiratory diseases, such as measles, pertussis, flu, and, since 2020, SARS-CoV-2, are
some of the key of public health concerns specifically due to their high viral potential.
This has made breaths and coughs, which are the most common symptoms among these airborne
diseases, the primary data in medical monitoring applications.

In the review, we focus mainly on summaries of the models and methods employed to achieve
the desired goal of automated cough analysis, detection, or disease classification.
The rationale is that, although most studies report remarkable sensitivity, specificity
and $F_1$ performance under different cross-validation approaches,
their use of private datasets, collected under differing acquisition protocols, or insufficient
reporting of performance on community-accepted benchmark datasets severely complicate comparisons.

The data in these datasets typically features spontaneous or induced coughs, obtained from
consenting participants that meet the requirements of a study and recorded on a dedicated
smart device. In detection studies, the audio is evaluated by the presence of ambient sounds
or speech, against which cough identification is to be performed. Disease classification
studies consider demographic factors, avoid collection site imbalances and curate the data
by the type and severity of the illnesses and recency of clinical tests for each particular
disease.
%
%

Prior to feature extraction and analysis, the audio signal is commonly preprocessed with
a low-pass filter with cut-off frequency at about $4$~kHz, the threshold determined in 
\citep{shin_automatic_2009} to contain most of the spectral content of a cough.
%
The respiratory data is then typically processed with the Short-time Fourier Transform,
i.e. the time-frequency representation computed by the discrete Fourier transform of
the signal in a sliding window with overlap. Subsequently, these spectrograms are transformed
using filter banks which prioritise higher resolution in lower frequencies, e.g. Mel-scale
(Mel-spectrogram) or Gammatone (cochleagram).
Although based on perceptually equal contribution to speech articulation, these filter banks
are frequently used for cough analysis since speech and cough generation share physiological
similarities, \citep{sharan_detecting_2021}.
In \citep{chatrzarrin_feature_2011}, it was observed that wet and dry coughs can be
distinguished by their spectral power distribution at \emph{initial burst}, \emph{noisy
airflow} and \emph{glottal closure} phases, specifically in the $1.5-2.5$~kHz and $\leq 750$~Hz
bands. This concentration of informative spectral content in lower frequencies justifies
the use of Mel-spectrograms and cochleagrams, \citep{sharan_detecting_2021}.

Research prior to 2017 tends to focus on \emph{acoustic or engineered features} -- the features
based on multi-scale wavelet, spectral or time-domain analysis commonly employed in signal
processing.
Examples include the rate of sign change in the time-domain, indices, and values of the local
extrema of the signal power in specific frequency bands, full or partial auto-correlation
coefficients, etc. However, the most often used features are the \emph{Mel-Frequency Cepstral
Coefficients} (MFCC), computed as the discrete cosine transform of the log-spectrogram aggregated
into Mel-scale frequency bins.
Analysis of the pairwise Mutual Information in \citep{drugman_assessment_2011} concluded
that the commonly used acoustic features complement each other in cough detection tasks.
Other engineered features used in the studies include spectral spread and centroid, Sample
Entropy, which measures the complexity and self-similarity of the time-series, non-Gaussianity
statistics, and Linear Predictive Coding features, %
\citep{matos_detection_2006,martinek_distinction_2008,ferdousi_cough_2015},
and \citep {pramono_cough-based_2016}.

These acoustic features are used in a large variety of cough detection methods: a keyword spotting
Hidden Markov model repurposed for continuous cough monitoring \citep{matos_detection_2006},
a cough-speech classification tree \citep{martinek_distinction_2008}, a 4-layer detector
network with additional bispectral features \citep{swarnkar_neural_2013}, three Hidden Markov
models stacked atop a neural network to distinguish cough from speech and ambient noise
\citep{shin_automatic_2009}, and a $k$-nearest neighbors classifier to detect the ``initial
burst'' phase of a cough, built on the PCA of acoustic features, \citep{lucio_voluntary_2014}.
In \citep{kochetov_noise_2018} respiratory cycle segmentation was done using a recurrent
network with an input noise masking mechanism driven by an auxiliary network, trained on
the MFCC features.
Going against the trend of increasing model complexity, the authors of \citep{pramono_automatic_2019}
insist on a minimalist approach to cough detection and show that a logistic regression
trained on just four acoustic features can deliver competitive sensitivity and specificity
metrics, comparable with the state of the art.

Methods for automated cough-based diagnosis of respiratory diseases borrow many ideas
and share technical similarities with detection.
For instance, pneumonia can be discerned from other respiratory infections with a logistic
regression on the acoustic features of a cough supplemented with wavelet transforms,
\citep{amrulloh_cough_2015,kosasih_wavelet_2015}, or pertussis can be identified by
a dedicated whoop detector and a cough classifier built with logistic regressions with
greedily selected features, \citep{pramono_cough-based_2016}. Chronic Obstructive Pulmonary
Disease can be identified with a random forest, \citep{windmon_detecting_2018}, or gradient
boosted classifier ensembles on the MFCC features, \citep{perna_automated_2017}. Finally,
the problem of croup diagnosis can be tackled with an SVM on 1\textsuperscript{st} and
3\textsuperscript{rd} moments of time-frequency sub-blocks of cough cochleagrams,
\citep{sharan_automatic_2019}.
Certain studies attempt to diagnose a condition without using coughs: a na{\"i}ve Bayes
classifier trained on acoustic features can detect if a patient has chronic cough purely
from the recordings of their speech, \citep{ferdousi_cough_2015}.

Since 2015 the cough detection and classification research has been gradually moving towards
\emph{deep features}, i.e. hierarchical features with different local receptive fields produced
by deep networks operating purely on spectrograms of the input signal or chunks thereof.
Two key motives prompted this shift. The first was the recognition that the commonly used
acoustic features were crafted for speech recognition applications, and, hence, inductively
biased towards the human auditory system and perception. The second was the wider adoption
of software frameworks for deep learning, coupled with greater availability of cutting-edge
deep networks pretrained on large datasets.

Many studies repurpose or fine-tune the state-of-the-art deep architectures to the tasks of
sound detection and respiratory disease classification. Indeed, it has been demonstrated that
hierarchical spectral features learnt by a convolutional neural network (CNN) and long-term
dependencies extracted by a recurrent network discriminate cough from speech and non-cough
sounds better than handcrafted acoustic features, \citep{amoh_deep_2016}. 
In \citep{bales_can_2020} a CNN operating on Mel-spectrograms is shown to successfully
diagnose bronchitis and bronchiolitis from the detected coughs.  
The study \citep{sharan_detecting_2021} identifies pertussis based on Mel-spectrograms and
cochleagrams of coughs fed into an ensemble of convolutional networks pooled by an SVM.
The author of \citep{saba_techniques_2018} adopts the paradigm of learning using privileged
information, and proposes an adversarial training method to suppress the effect of undesirable
confounding variables on the outcomes of a convolutional cough-based tuberculosis classifier.
The study \citep{barata_towards_2019} addresses the issue of varying quality of recordings
by developing a device-agnostic bagging ensemble of architectures inspired by VGG-19,
\citep{simonyan_very_2015}.
%

With the onset of the SARS-CoV-2 pandemic the volume of research on the feasibility and public health
capabilities of fast and inexpensive audio-based diagnostics of respiratory illnesses have
grown considerably, \citep{huazhu_fu_covid-19_2021}. In particular, lung X-Ray and CT scan
studies suggest that COVID-related coughs should have idiosyncratic signatures stemming from
distinct underlying pathomorhpology, \citep{imran_ai4covid-19_2020}.
%
%
Surveys \citep{asraf_deep_2020} and \citep{ulhaq_covid-19_2020} provide references to non
audio-based deep learning solution related to COVID, including disease identification from
medical images and ultrasound, contact and spread tracking and tracing using facial recognition,
screening of respiratory patterns, and protein analysis for drug discovery and virulence
prediction.

%
Studies using respiratory audio related to COVID-19 collect curated datasets of sounds
and symptoms from local hospitals or wards, paying special attention to patient eligibility
criteria, prior knowledge of COVID status, and imbalance due to acquisition time, location,
demographics, the equipment, and the hardware used to capture and record audio,
\citep{pinkas_sars-cov-2_2020,bagad_cough_2020,han_early_2020,andreu-perez_generic_2021}.
%
%
Other studies crowdsource the data through web or mobile apps, which is a more affordable
and less time-consuming option, that yields much larger datasets, albeit of lesser quality
both in the ground truth infection status labels and the audio recordings themselves,
\citep{brown_exploring_2020,orlandic_coughvid_2020,laguarta_covid-19_2020}.
%

The machine learning methods and pipelines considered for cough-based COVID screening
have for the most part continued prior audio-based disease identification research.
%
For example, \citep{brown_exploring_2020} identify COVID with non-deep classifiers
trained
on a subset of their crowdsourced dataset using acoustic features and \emph{VGGish}
embeddings \citep{hershey_cnn_2017} of breath cycles and coughs.
A preliminary study \citep{han_early_2020} uses an extended acoustic feature set in
a class-balanced linear support vector classifier to predict sleep quality, fatigue,
anxiety levels and a proxy for COVID severity on $51$ patients from a COVID ward
without a non-COVID control group.
%
In \citep{andreu-perez_generic_2021} the authors collect a dataset of $\approx8300$
cough samples from patients with clinically verified qRT-PCR test outcome on which
they develop a deep detection and COVID infection severity classification system,
that operates on the Mel-spectrogram, the MFCC, and sliding partial autocorrelations.
%
Another study \citep{imran_ai4covid-19_2020} develops a screening tool that vets
the input audio for coughs and combines the outputs of a committee of intermediate
heterogeneous classifiers into a final COVID diagnosis by unanimous voting, abstaining
in case of discord. SVM classifiers in the ensemble are built on the aggregated
MFCC features and their principal projections, while the CNN classifiers are trained
on Mel-spectrograms.

The diagnostic model from \citep{laguarta_covid-19_2020} aggregates salient information
from biomarkers computed from MFCC of the input audio with three ResNet-50 models,
\citep{he_deep_2016},
independently fine-tuned to detect sentiment,
to measure vocal cord fatigue,
and to capture acoustic idiosyncrasies due to respiratory tract structure.
%
%
In another recent study \citep{bagad_cough_2020}, the authors build a two-layer classifier
atop the deep feature extractor of a pre-trained ResNet-18, \citep{he_deep_2016}. Their
model is then fine-tuned on Mel-spectrograms for a cough detection task on a pooled
non-COVID dataset of speech and respiratory
sounds, \citep{fonseca_general-purpose_2018,al_hossain_flusense_2020,sharma_coswara_2020},
with each
sample contaminated by a random background environmental noise, \citep{piczak_environmental_2015}.
Finally, the model is further fine-tuned to identify COVID-positive coughs from a carefully
curated dataset of three thousand samples collected from testing sites and COVID wards in
India.
%
The ablation study with stratified grouped cross-validation in \citep{bagad_cough_2020}
demonstrates that cough-non-cough pretraining and ensembling contribute positively to
the performance of a stacked composite COVID-classifier, consisting of the deep ResNet-18
classifier and shallow models on acoustic features from \citep{brown_exploring_2020}.
The approach in \citep{pinkas_sars-cov-2_2020} departs from cough event analysis and instead
repurposes the voice embeddings, produced by a pre-trained transformer speech model, in
an SVM-stacked ensemble of deep recurrent classifiers trained on $292$ voice samples of $88$
patients with COVID status verified by the RT-PCR test.

\section{Datasets}
\label{sec:datasets}

Datasets that were used in this work can be split into three types: \emph{prior crowdsourced},
\emph{curated clinical}, and \emph{newly collected} data. Namely:
\begin{enumerate}
    \item Openly accessible moderately large \emph{crowdsourced cough datasets} with unverified
    labels collected by other projects and researchers, e.g. \citep{fonseca_learning_2019},
    \item Smaller size \emph{curated private datasets} with verified labels from hospitals and
    COVID wards in Russia acquired under proper participation and recording protocols,
    \item \emph{New cough and symptom data}, being continuously collected in uncontrolled
    conditions from users of our mobile application since its initial release
\end{enumerate}
This section describes all these types of data in more detail.

\subsection{Open Datasets} 
\label{sub:open_datasets}

As of the time of writing, only three large cough datasets featuring COVID-19 positive
samples were publicly available -- the EPFL COUGHVID dataset~\citep{orlandic_coughvid_2020},
Coswara~\citep{sharma_coswara_2020}, and Covid19-Cough~\citep{project_fkthecovid_dataset_2021}.

The EPFL dataset comprises $20072$ records with $1010$ self-reported COVID-positive.
Each record contains cough recording and additional metadata like age, gender, symptoms,
geographical location. Part of the dataset is annotated by three doctors.
The Coswara dataset consists of about $2000$ records including about $400$ from positive
patients. Similarly to COUGHVID it contains additional metadata and the COVID status
is self-reported.
The Covid19-Cough dataset consists of $1324$ samples with $682$ COVID-positive cases,
$382$ of them confirmed by a PCR test. Samples were collected through a call-center and
via Telegram messenger bot (see Table \ref{tab:russian_data_table}).
%

\begin{table}[!t]
\caption{
    The label distribution within different slices of the \emph{Covid19-Cough} dataset
}
\label{tab:russian_data_table}
\centering
\resizebox{0.4\textwidth}{!}{%
\begin{tabular}{@{}lrr@{}}
    \toprule
        & \multicolumn{2}{c}{\textbf{COVID status}} \\
    \midrule
        & \textbf{Positive} & \textbf{Negative} \\
    \cmidrule(l){2-3} 
    \multicolumn{1}{l}{\textbf{Telegram source}}
        & 186 & 438 \\
    \multicolumn{1}{l}{\textbf{Call-center source}}
        & 496 & 204 \\
    \multicolumn{1}{l}{\textbf{Total}} 
        & 682 & 642 \\
    \multicolumn{1}{l}{\textbf{COVID asymptomatic}} 
        & 379 & --- \\
    \multicolumn{1}{l}{\textbf{Verified diagnosis}} 
        & 382 & 0 \\
    \bottomrule
\end{tabular}%
}
\end{table}

\subsection{Proprietary Data} 
\label{sub:private_data}

As mentioned earlier, the affordability of the crowdsourcing option comes at the cost
of control over selection bias, confounding variables, and reliability of the ground-truth
labels. The open datasets in the previous section suffer from these shortcomings,
especially since the COVID status labels are self-reported and mostly unverified by
a PCR test, i.e. \emph{weak}.
%
In order to make a mobile application using a deep learning model, which is capable
of detecting COVID on cough, breath, and speech input, we have collected a private
dataset with the infection status, verified by a PCR test. We expect that fine-tuning
on a strongly labeled cough dataset would reduce the potential classification bias of
the model trained on abundant, but weaker data, and, therefore, improve the final performance.
%

Each sample in the collected dataset consists of three audio recordings, symptom data,
and COVID status. Every person who agreed to participate in the study has been recorded
only once. We make sure each participant gives informed explicit consent prior to uploading
their respiratory and voice samples and medical data to a cloud data storage for later
processing.
%
The type of collected data is similar to \citep{brown_exploring_2020}, but rather than
asking for a specific number of isolated respiratory events, we limit the duration of
of the recorded continuous coughing and breathing.
%
We obtain two five-second audio samples of induced cough and breathing cycles, and
a recording of the vocalized recitation of the Russian phrase
\emph{``I hope, this recording will help battle the pandemic''}.

Audio samples for COVID-positive cases were collected in hospitals and verified by both
a PCR test and lung CT scan. COVID-negative participants were recorded in an office
environment which required a recent and verified negative PCR test for entry.
Ultimately, we obtained $211$ samples from healthy users and $228$ records from COVID-19
wards from two hospitals in Moscow, Russia. Although these samples could bias the dataset
towards severe COVID-positive cases and provide little help in detecting asymptomatic
carriers, when used to fine-tune the model they appear to improve its performance
(see sec.~\ref{sec:experiments_and_discussion}).
Since the number of samples in our private dataset was limited we have decided not to
allocate a test set for testing the model's performance, but keep a comparatively small
subset of the collected data mainly for inevitable unit tests.
This subsample comprises of $17$ randomly chosen recordings from COVID wards and eight
random recordings of healthy people.
We have one exclusion from this approach, namely, we used this small held out dataset
to combine distinct ensembles into a single meta-ensemble (section \ref{sub:stacking_and_results_on_app_data}.
The rest of the collected private data was pooled with the Covid19-Cough dataset to
fine-tune our models (section \ref{sub:finetuning}).
%
%

Along with the audio data, we have collected self-reported subjective symptom data by
asking the participants to pick the symptoms from a list in Table~\ref{tbl:logreg_stats},
which correlates with prior medical or clinical studies, \citep{grant_prevalence_2020} and
\citep[sec. I.C, II.A, and~II.B]{imran_ai4covid-19_2020} and was approved by practicing
medical experts. The case numbers in the table reflect the incidence rate in the collected
dataset.
%

\subsection{App data}
\label{sub:app_data}

The trained and validated models were deployed as a backend of our custom iOS~/~Android
application, which we have used to collect similar respiratory audio, speech, and symptom
data.
No personal data or metadata was collected, other than the device model information,
useful for adjusting for possible bias associated with the hardware.
It is worth noting that this dataset is prone to label noise since COVID status is
self-reported by the users, in contrast to the \emph{strongly} labeled dataset, mentioned
in the previous section ( the ``Hospital'' dataset).

We collected the validation dataset for section~\ref{sec:application} and Figure~\ref{fig:all_datasets}
(``app data'') on the first day after the release of the application.
%
During this time we collected $1395$ records from $1035$ unique devices: $901$ data
points from $700$ unique iOS devices and $494$ -- from $335$ unique Android devices.
%

\begin{table}[!t]
\caption{
    The most salient symptoms for COVID diagnosis.
}
\label{tbl:logreg_stats}
\centering
    \begin{tabular}{lr}
    \toprule
        Symptom          & Cases \\
    \midrule
        Diarrhoea        &   9   \\
        Dyspnoea         &  39   \\
        Sore throat      &   7   \\
        Cough            &  62   \\
        Rash             &   4   \\
        Fatigue          &  83   \\
        Fever            &  19   \\
        Anosmia          &  41   \\
        Dry tongue       &  27   \\
    \midrule
        COVID positive   & 102   \\
        Total cases      & 154   \\
    \bottomrule
    \end{tabular}
\end{table}

\section{the Method}  
\label{sec:the_method}

Ensemble methods combine weak predictors into composite models with reduced bias or
variance with the goal of improving prediction performance, \citep{hastie2008elements}.
Stacking uses a trained meta-model to combine raw predictive outputs of independent
intermediate models, and bagging averages predictions of separate unbiased predictors
decorrelated by data- and feature-level bootstrapping, random projections, and other
methods.
Boosting builds a superior predictive model by blending intermediate weaker ones through
what amounts to stochastic gradient descent, with each model estimating a finite-sample
approximation of the functional derivative of the loss.

\begin{figure}[!t]
  \centering
  \includegraphics[width=0.85\columnwidth]{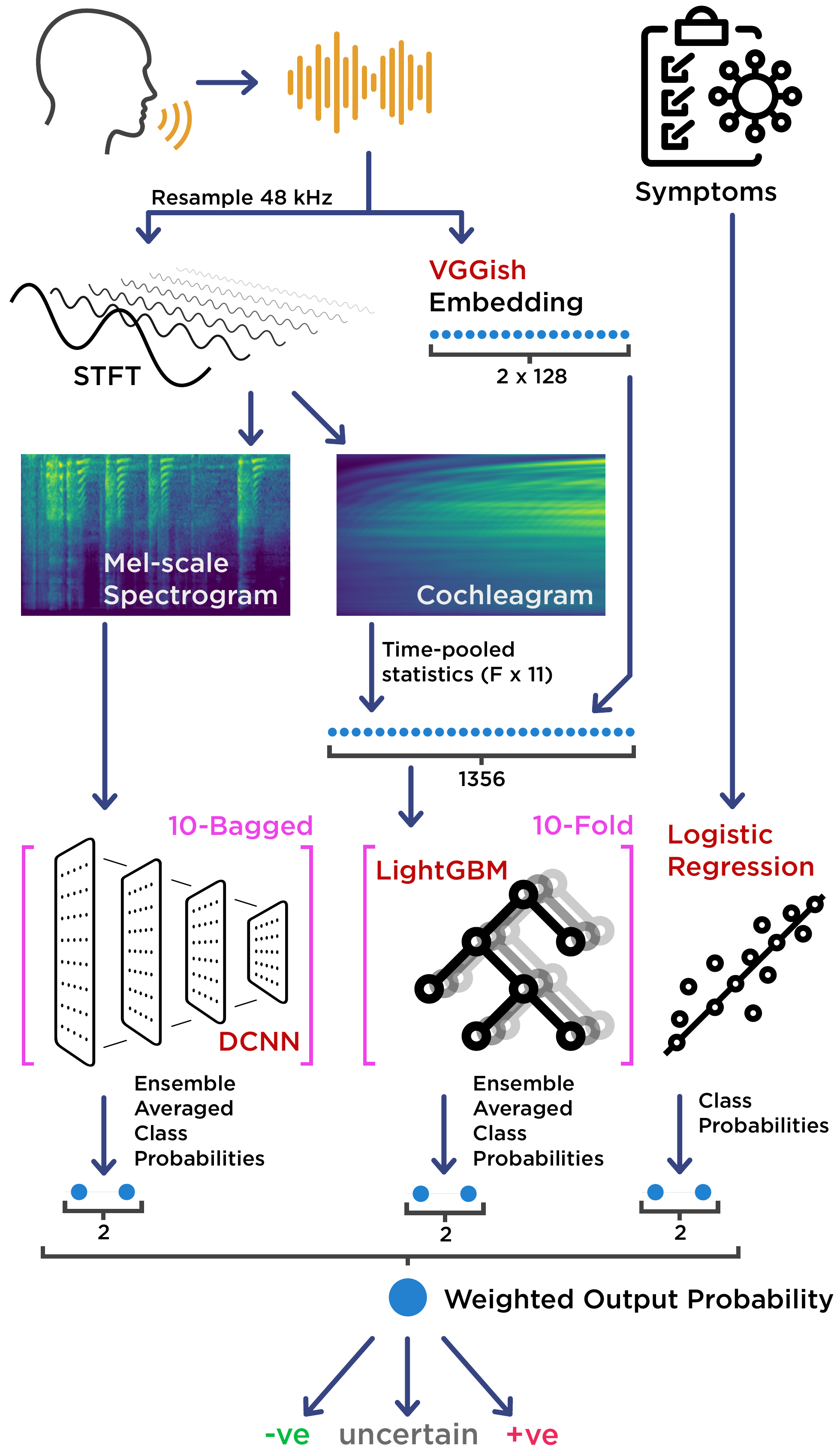}
  \caption{The diagram of the entire pipeline of our solution.}
  \label{fig:deep_branches_of_the_hybrid}
\end{figure}

%
The method we use in this study employs a hybrid ensemble approach (Figure~\ref{fig:deep_branches_of_the_hybrid}) 
-- we bag heterogeneous classifiers trained and fine-tuned on multiple datasets and stack
a second level meta-model. It is possible to assign different weights to the intermediate
models in order to achieve the desired trade-off in the classification performance metrics
(Table~\ref{tab:ensembling}).
The overall ensemble analyzes the input respiratory audio using a diverse set of learnt
and tuned patterns and supplements the detected ``signal'' with simple-to-observe, yet
informative symptom data. By carefully combining fine-tuned predictors in the ensemble
we improve the overall prediction quality of our method and achieve favorable bias-variance
trade-off. The pipeline, depicted in Figure \ref{fig:deep_branches_of_the_hybrid}, represents
the workflow of the app (see Section \ref{sec:application}).
During training, each model in the ensemble outputs a probability of a binary label, without
considering the possibility of ``abstention'' or ``indecision''.
%

The pipeline starts with the quality control step, which uses a deep detector to filter
out audio recordings that do not contain cough events (sec.~\ref{sub:cough_detection}).
At the preprocessing stage, we extract the VGGish features from the audio sample, \citep{hershey_cnn_2017},
and compute its Mel-scale spectrogram representation (sec.~\ref{sub:spectrogram}) as
well as the time-aggregated statistics of its cochleagram (sec.~\ref{sub:spectrogram}).
The spectrogram is fed as-is into a bagged ensemble of deep convolutional networks trained
and fine-tuned on sub-sampled datasets. At the same time, VGGish features are combined with
the cochleagram statistics and passed into the gradient boosting ensemble.
In parallel to the ensemble models, there is a binary COVID classifier trained on user-reported
flu-like symptoms.
%
The final step is computing the weighted average of the intermediate classifiers' output
probabilities.


In the remainder of this section, we detail our approach to selecting the preprocessing
parameters, architectures and hyperparameters of the classifiers and ensembles.

\subsection{Mel-spectrogram}  
\label{sub:spectrogram}

For the deep convolutional models, all audio recordings were resampled to $48$~kHz and
the leading and trailing silence was trimmed. We did not apply any frequency filtering
in order to preserve as much spectral data as possible. Afterwards, the cough waveforms
were converted into time-frequency representation by the Short-Time Fourier Transform over
sliding $54$~ms frames with $14$~ms strides and Hann windowing function using \emph{librosa}
package. The Mel-spectrograms were obtained by projecting the representations
into $128$-bin Mel filter-bank spanning $20$~Hz~-~$24$~kHz frequency range.
During training, the Mel-spectrograms were augmented by randomly cropping or replicating
them along the time axis to get same-duration chunks of roughly eight seconds.
We also introduce auxiliary frequency-bin positional encoding as the second feature channel
of the spectrogram.



We used the cochleagram statistics aggregated across the temporal dimension as the inputs
for the gradient boosting ensemble model in our pipeline (Figure~\ref{fig:deep_branches_of_the_hybrid}).
These features were extracted for cough samples, and, if available, from breath and voice
recordings.
The input signal was transformed into cochleagrams using \emph{Brian2Hears} package\footnote{
    Auditory modelling toolbox. \url{https://github.com/brian-team/brian2hears}
}, with the number of frequency bins set to $100$ and other parameters kept at their default
values.
%
%
Next, we obtained a time-frequency representation of the input signal as a matrix with $100$
rows, one for each bin in the cochleagram, and $n_c$ columns, the number of which is determined
by the duration of the input.

For each frequency bin, i.e. the vector of dimension $n_c$, we computed $11$ values:
the mean, median, standard deviation, skew, kurtosis, minimum, maximum, the first $Q_1$
and the third $Q_3$ quartiles, the interquartile range ($Q_3 - Q_1$) and $\ell_2$-norm.
The resulting feature matrix $100 \times 11$ was flattened and joined with the input's VGGish
$256$-dimensional embeddings, \citep{hershey_cnn_2017}. Ultimately, we obtained a feature vector
of length $1356$, that served as input into the subsequent ensemble
(Figure~\ref{fig:deep_branches_of_the_hybrid})




%

\section{Experiments and discussion} 
\label{sec:experiments_and_discussion}

\begin{figure}
    \centering
    \includegraphics[width=0.9\columnwidth]{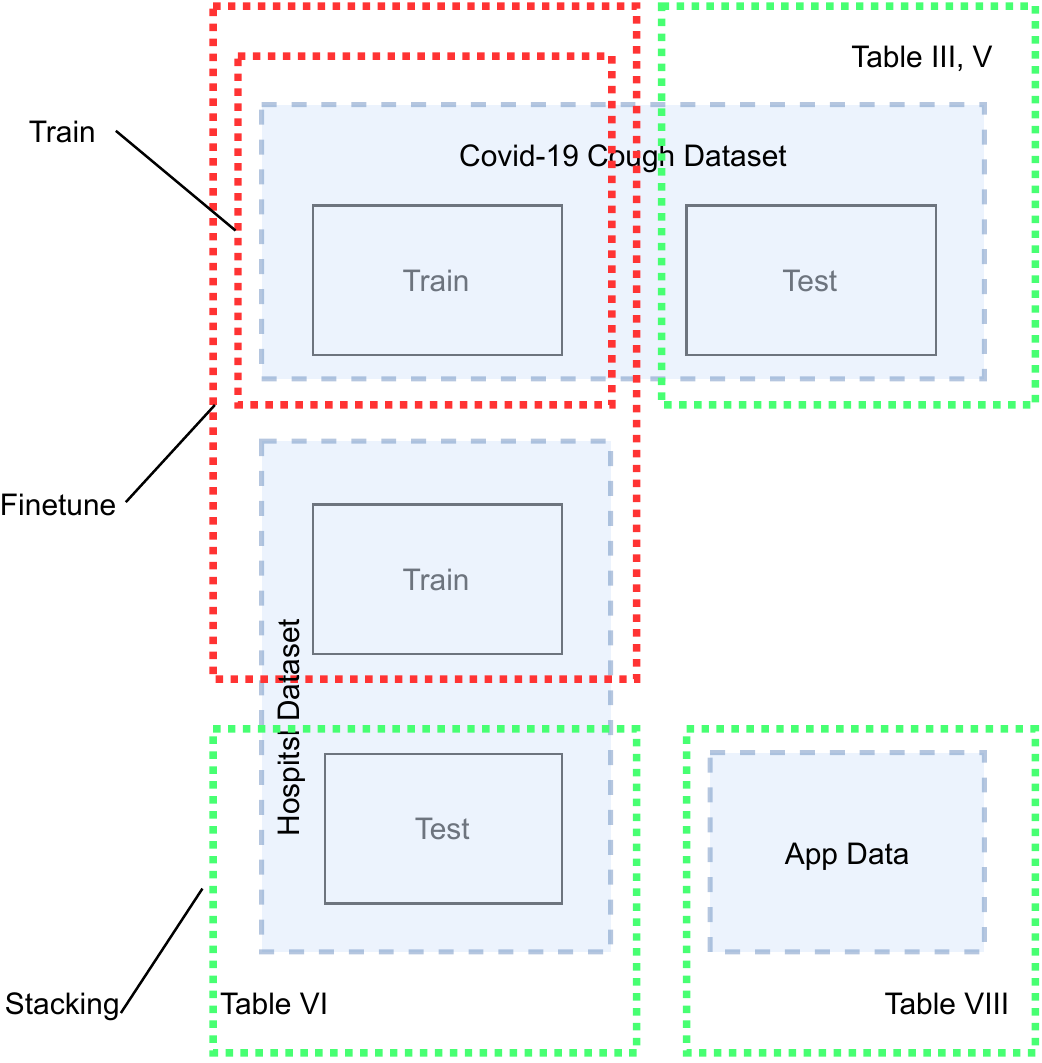}
    \caption{
        Datasets used in our work.
        Models were trained on the Covid19-Cough dataset and fine-tuned on the Covid19-Cough
        and the part of the ``Hospital'' dataset.
        {\color{red}{Red}} dotted boxes indicate the data that was used in some way to change
        the weights of our models.
        {\color{green}{Green}} dotted boxes indicate those parts of datasets that were used
        for stacking and/or calculating the target metrics.
    }
    \label{fig:all_datasets}
\end{figure}

In this section we give a brief description of our training and fine-tuning procedure
and introduce datasets we used (Figure \ref{fig:all_datasets}); in the following section, we dive into details.
\subsection{Training Procedure Overview}

%
At the first stage we chose the best performing publicly available dataset, \emph{Covid19-Cough} (see
sec.~\ref{sub:open_datasets} and Table \ref{tab:res_on_open_data}) and further used it
to train our models: deep CNN and gradient boosting. 
We then collected two datasets: a dataset with ``strong'' labels, i.e.
samples with comparatively good record quality and verified labels (we also refer to
this dataset as ``Hospital'' dataset, see sec.~\ref{sub:private_data}); and a crowdsourced dataset (App Data), that was collected via our app (see \ref{sub:app_data}).
The former was split into a training set and a held out test set.
%
The training set was used to fine-tune models fit on the open Covid19-Cough dataset,
while the test subset was used i) for software unit testing and ii) to select parameters
for stacking models into the ensemble (see sec.~\ref{sub:finetuning}).
We evaluate i) the classification quality of fine-tuned models on the Covid19-Cough
dataset and ii) ensembles, obtained after stacking models on the test dataset and
on the crowdsourced data collected via the app (see Table \ref{tab:performance_on_app_data}).

\subsection{Architecture and Fine-tuning}
\label{sub:acoustic_arch_finetune}

We used a modified light-weight CNN architecture based on \citep{fonseca_learning_2019},
that supplemented the Mel-spectrogram image-like input by an additional channel corresponding
to log-frequency positional encoding. The extra channel propagates through skip connections
into deeper layers, which enables better localization and utilization of frequency information.
On input the convolutional model receives the recorded sound of coughs preprocessed as
described in sec.~\ref{sub:spectrogram}.

In order to train an ensemble of DCNN and a bagged ensemble of Gradient Boosted Trees we
utilized $10$-fold cross-validation on openly available datasets.
Each fold was further split into train and validation subsets, which altogether
correspond to a random $70-15-15$ train-validation-test split of the dataset.

We trained the gradient boosted classifier ensemble with LightGBM \citep{ke_lightgbm_2017}
using identical hyper-parameters on each fold, that were fine-tuned on validation datasets.
The maximum number of leaves in a tree was $5$, while the minimum number of data samples in
one leaf was $35$. We set the learning rate to $1\cdot {10}^{-1}$ and optimized the weighted
cross entropy loss with $\ell_2$ regularization coefficient ${10}^{-3}$.
The models were further fine-tuned using our private ``Hospital'' dataset (sec.~\ref{sub:private_data}).

\begin{figure*}
\begin{multicols}{2}
    \includegraphics[width=\linewidth]{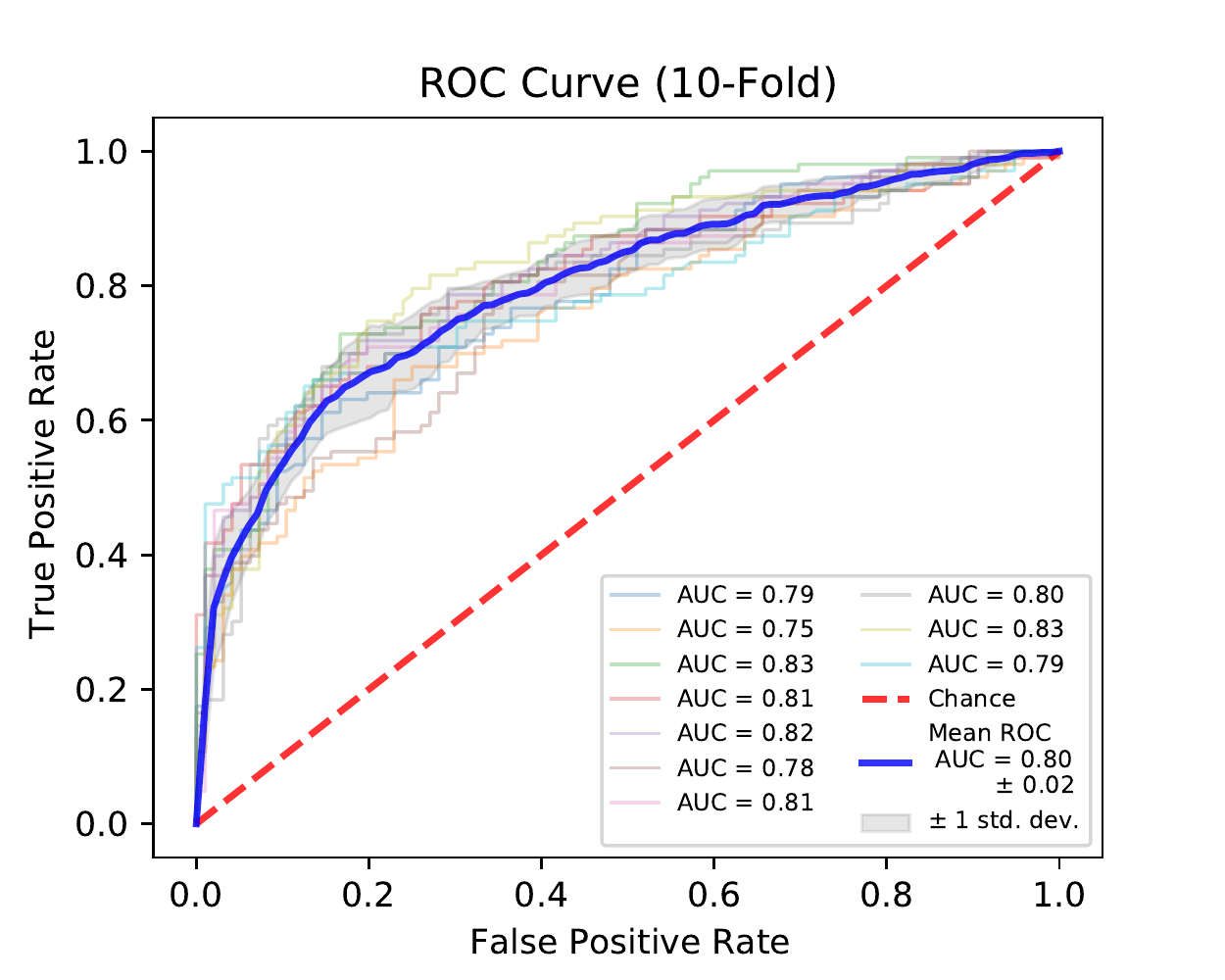}\par 
    \includegraphics[width=\linewidth]{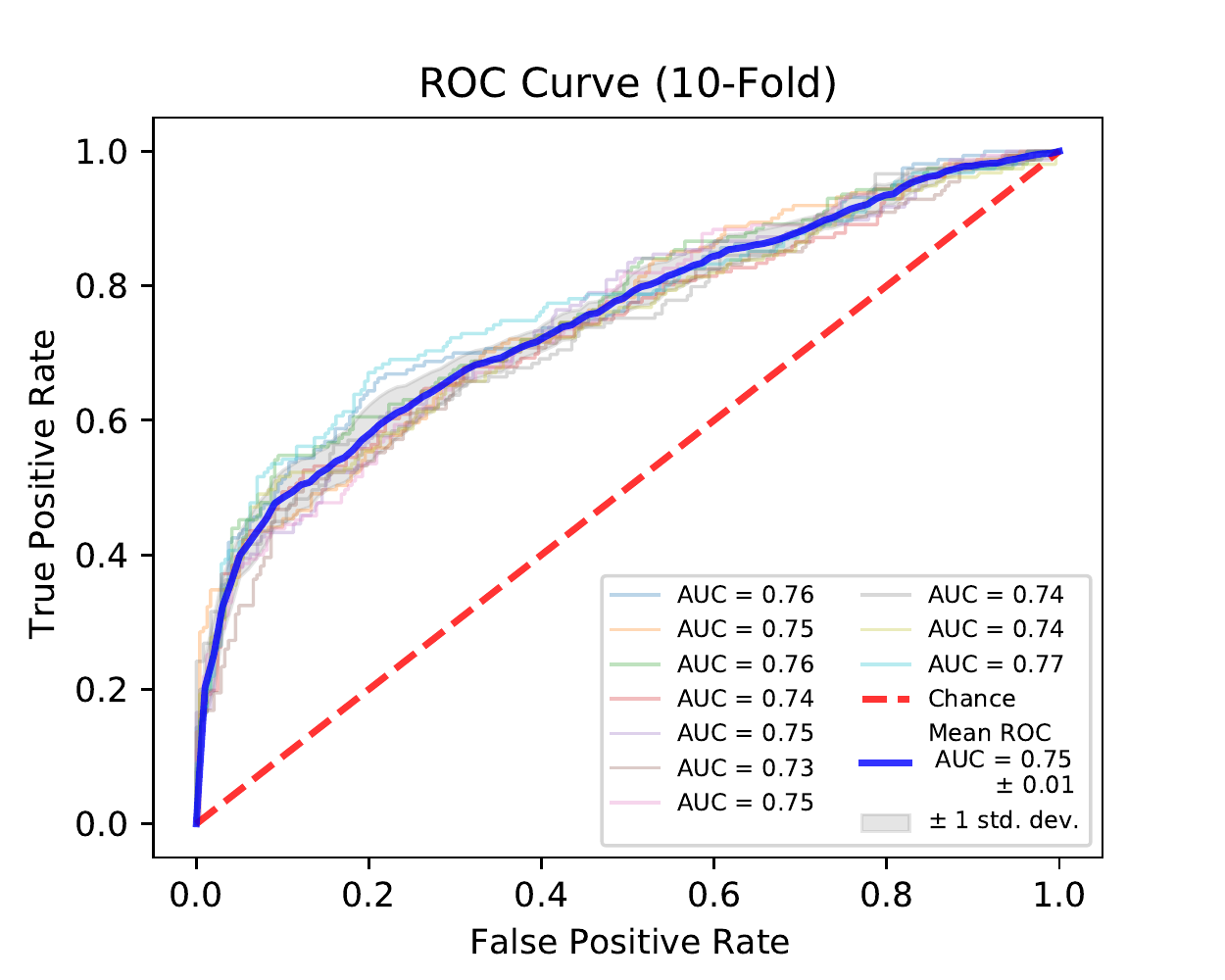}\par 
    \end{multicols}
    \caption{
        The ROC curves for predictions of models on $10$ distinct test datasets from Covid19-cough.
        Blue depicts the mean ROC curve calculated by averaging false positive rates
        and true positive rates of individual predictors.
        The ROC curve on the left was built for deep neural networks; on the right --
        for gradient boosting.
    }
\label{fig:roc_curves_test}
\end{figure*}

Figure \ref{fig:roc_curves_test} depicts ten ROC curves for each test split in the $10$-fold
cross-validation, with the average ROC AUC value of $0.7473$. 
We also tried to adopt the same training strategy for records of breath and speech, however, these ensembles did not substantially improve overall performance. 

\subsection{Performance on open datasets}
\label{sub:perfromance_open_data}

In order to rank the open-access datasets in terms of the label quality we employed
the following ad-hoc approach. We scored each openly accessible crowdsourced
dataset by its $10$-fold averaged ROC AUC and Matthews Correlation Score (MCC),
\citep{chicco_advantages_2020}, independently computed using each branch of our
model (sec.~\ref{sec:the_method}, Figure~\ref{fig:deep_branches_of_the_hybrid}).
Each replication in the $k$-fold CV was split into 70\%-15\%-15\% for train,
model-selection, and test subsets, respectively.

\begin{table}[!t]
    \caption{Classification metrics of different predictors on openly available crowdsourced datasets}
    \label{tab:res_on_open_data}
    \centering
    \begin{tabular}{l|r|r|r|r}
    \toprule
        \multirow{2}{*}{Dataset}         &  \multicolumn{2}{c|}{ROC AUC}  & \multicolumn{2}{c}{MCC} \\
        \cmidrule{2-5}
        & CNN & GB & CNN & GB \\
    \midrule
        EPFL            & 0.630 & 0.592 & 0.112 & 0.154 \\
        Coswara         & 0.705 & 0.743 & 0.219 & 0.073 \\
        Covid19-Cough   & \textbf{0.801} & 0.747 & 0.456 & 0.408\\
    \bottomrule
    \end{tabular}
\end{table}

The computed cross-validated classification scores are presented in
Table~\ref{tab:res_on_open_data}.
When measured with the CNN branch, Coswara and the EPFL datasets exhibit more
or less the same quality close to random guessing, while the Covid19-Cough dataset
scores tangibly higher. Since the convolutional model used to get the scores is
not over-parametrized to memorize the dataset, \citep{ovadia2019canyou}, and its
architecture was shown to be effective in applications (\citep{fonseca_learning_2019},
and Kaggle Freesound competition\footnote{
    \url{https://www.kaggle.com/c/freesound-audio-tagging-2019}
}), we speculated that the variability of ROC AUC and the MCC scores between
the datasets could be due to the potentially mislabelled COVID status in Coswara
and the EPFL datasets, which, unlike Covid19-Cough, are also highly imbalanced.
To evaluate this conjecture, we investigated the subset of recordings in
the EPFL dataset that were additionally assessed by practicing medical doctors,
\citep{orlandic_coughvid_2020}. Despite adequate recording quality indicated
by the experts, these labels appeared to be uncorrelated with each other and
with self-reported COVID status, which further lent evidence to the presence
of COVID status noise and could explain the apparent disparity in ROC AUC and
the MCC scores (Table~\ref{tab:inter_expert_mcc_epfl}).
%

\begin{table}[t!]
    \caption{
        The MCC scores calculated for experts' predictions and self-reported labels
        on the EPFL dataset.
    }
    \centering
    \begin{tabular}{l|rrrr}
    
    \toprule
    \backslashbox{Pred.}{True}
        & Self-reported & Expert 1 & Expert 2 & Expert 3 \\
        \hline
        Self-reported & 1.     & -0.051 & -0.036 & -0.024\\
        Expert 1      & -0.051 & 1.     & 0.033  & 0. \\
        Expert 2      & -0.036 & 0.033  & 1.     & 0. \\
        Expert 3      & -0.024 & 0.     & 0.     & 1. \\
    \bottomrule
    \end{tabular}
    \label{tab:inter_expert_mcc_epfl}
\end{table}

\subsection{Fine-tuning}
\label{sub:finetuning}
 
Models trained on the Covid19-Cough dataset were further fine-tuned using our ``Hospital'' data (see \ref{sub:private_data}).
We fine-tuned the ensemble of CNN using records of cough.
The gradient boosting ensemble was fine-tuned using records of cough, breath
and vocalization, which we call GB, GB\textsubscript{b} and GB\textsubscript{v}
respectively, resulting in three ensembles from Table \ref{tab:finetune_models}.
The private data that we collected in hospitals was split into ten folds and combined fold-wise with
the Covid19-Cough dataset.
The goal of this stage was to train our models on data that are close to those that will be recorded
by the users of our final application.

We report the mean classification metrics before and after fine-tuning, respectively, in columns
``Covid19-Cough'' and ``Joined training set'' of Table \ref{tab:finetune_models}.
For both ensembles, CNN and GB, we provide mean ROC AUC and the MCC metrics averaged over
the ten held out test subsets of the Covid19-Cough dataset. This explains why the MCC decreases
while ROC AUC slightly increases for the gradient boosting.
%

\begin{table}[!t]
    \caption{
        The average model performance for $10$-fold cross-validation with Covid19-Cough dataset and
        joined Covid19-Cough and Hospital dataset.
        Testing was performed on the same test parts of CV splits of the Covid19-Cough dataset.
    }
    \label{tab:finetune_models}
    \centering
    \begin{tabular}{l|r|r|r|r}
    \toprule
        \multirow{2}{*}{Model}
            & \multicolumn{2}{c|}{Covid19-Cough} 
                & \multicolumn{2}{c}{Joined training set} \\ 
    \cmidrule{2-5}
            & ROC AUC & MCC
                & ROC AUC & MCC \\
    \midrule
        CNN       & 0.801    & ---    & \textbf{0.805} & --- \\
        GB        & \multirow{3}{*}{0.747} & \multirow{3}{*}{0.408} & 0.749 & 0.385 \\
        $\text{GB}_b$&&&0.749&0.364\\
        $\text{GB}_v$&&&0.686&0.383\\
    \bottomrule
    \end{tabular}
\end{table}

\begin{table*}[t]
    \caption{
        Distinct variants of ensembling lead to different performance on the test dataset.
        We are able to change weights in the ensemble in order to maximize specific metrics.
    }
    \label{tab:ensembling}
    \centering
    \begin{tabular}{l|r|r|r|r|r|r}
    \toprule
    Name
        & $t$ & $x$ & $y$ & $z$
            & ROC AUC &  MCC\\    
    \midrule
    Ensemble (Variant I)
        & 0.02 & 0.412 & 0.284 & 0.284 
            & \textbf{0.956} & 0.618\\
    Ensemble (Variant II)
        & 0.20 & 0.656  & 0.   & 0.144  
            & 0.897 & \textbf{0.736} \\
    \bottomrule
    \end{tabular}
\end{table*}


\subsection{Stacking and Results on data from the app}
\label{sub:stacking_and_results_on_app_data}

%
We stacked the four previously trained ensembles (three gradient boosting ensembles from the previous section and the ensemble of CNN)
using grid search over weights assigned to each ensemble's prediction to maximize the classification
metrics on the test data of our private dataset (see sec.~\ref{sub:private_data}). 

The goal of this stage was to find optimal weights that would maximize our target metrics while
using the crowdsourced data from the app.
We realize that the mentioned test dataset is relatively small, but we aimed to determine whether
a recording of breath and vocalization would be able to improve the ensemble's performance on
crowdsourced data.
In Table \ref{tab:ensembling} we report two variants of combining fine-tuned ensembles that
maximize certain classification metrics. 
Every sub-ensemble output probability averaged over $10$ distinct predictors, thus the final
probability of an ensemble is given by the formula:
\begin{equation*}
p   = t \cdot p_{\mathrm{DCNN}}
    + x \cdot p_{\mathrm{GB}}
    + y \cdot p_{\mathrm{GB_b}}
    + z \cdot p_{\mathrm{GB_v}}
    \,,
\end{equation*}
where GB, GB\textsubscript{b}, and GB\textsubscript{v} stand for the gradient boosted ensemble fine-tuned, respectively, on
cough, breath, and vocalization data.

Next, we provide the performance of our models on App data (see sec.~\ref{sub:app_data}) in Table \ref{tab:performance_on_app_data}.
Results of CNN, GB, and Ensembles (Variants I and II) were measured using recordings of coughs,
while GB-breath and GB-vocalization were measured on breath and vocalization data respectively.
Ensembles fine-tuned on breath and vocalization have no predictive power per se, while ensembles
trained and fine-tuned on coughs are more successful.
This raises the question of the possibility of using such recordings for achieving good results
in limited computational time.
Another indication that cough recordings have higher signal compared to breath and
vocalization is that among the two variants of our stacked ensemble, the best performance is achieved
by a configuration with higher weights assigned to cough sub-ensembles, namely CNN and GB.
Note that the results of CNN, GB are on par with those of Ensemble (Variant II).

\begin{figure}[!t]
    \centering
    \includegraphics[width=0.45\textwidth]{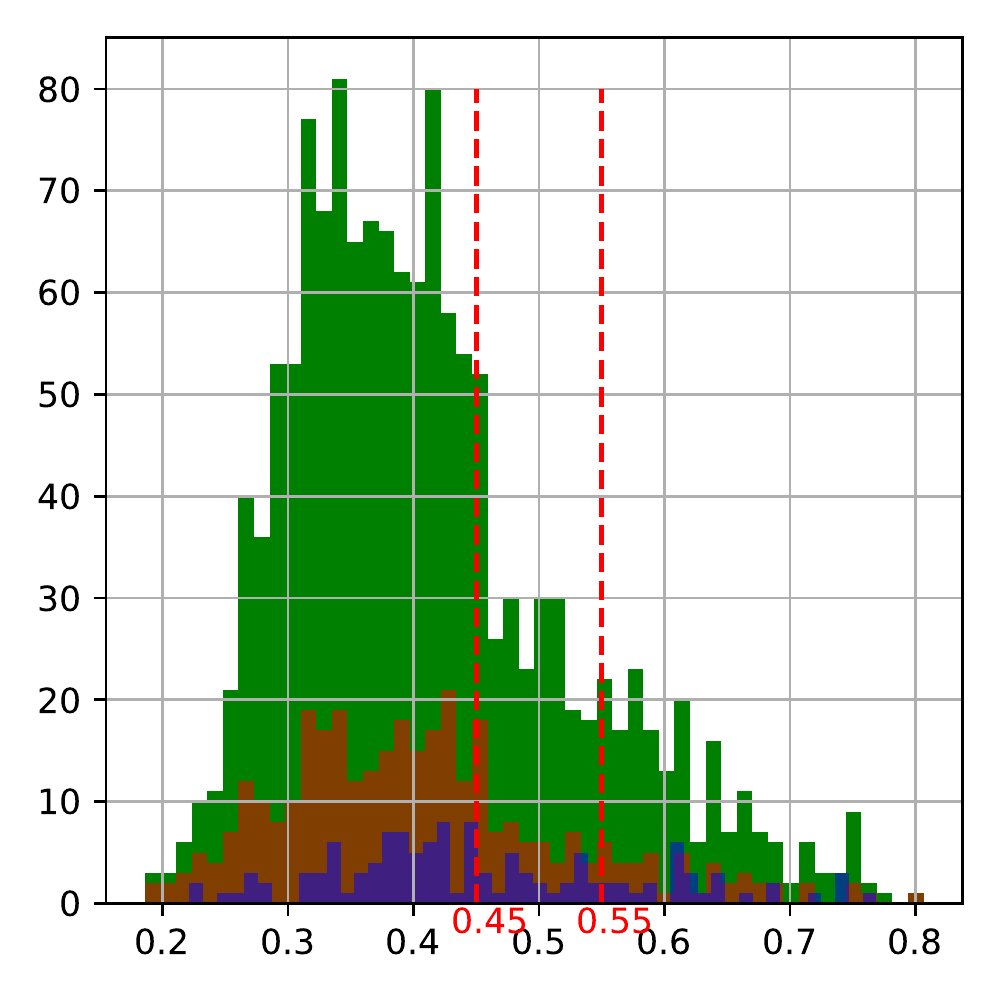}
    \caption{
        Histogram depicts how output probabilities of the Ensemble (Variant II) are distributed.
        In green -- all cases; in red -- cases where users reports absence of a respiratory disease;
        in blue -- cases where users reports presence of a respiratory disease.
    }
    \label{fig:prob_dest_reall_data}
\end{figure}

Not all users reported their health status -- whether they are afflicted by any acute respiratory
disease at the moment of recording. 
In Figure \ref{fig:prob_dest_reall_data} we summarised how probabilities of one of our ensembles
(Variant II) are distributed. 

{\color{green}{Green}} bars correspond to the overall distribution, {\color{red}{red}} bars
correspond to those cases where people verified that they do not have any acute respiratory
diseases at the moment of recording, and {\color{blue}{blue}} bars correspond to those people
who confirmed that they have respiratory disease(s) at the moment of recording.
The {\color{red}{red}} dotted lines correspond to uncertainty thresholds mentioned in
section \ref{sec:the_method}, i.e. if the model predicts presence of the COVID with probability
$0.45 \leq p(x) \leq 0.55$, the app informs the user about the model's uncertainty,
and allows the user to repeat the process from the beginning.

We see that the histogram of probabilities (Figure~\ref{fig:prob_dest_reall_data}) is biased towards lower probabilities, namely,
80\% of outcomes have probability lower than $0.5$. For those cases where people indicated
\emph{absence} of acute respiratory diseases, $p=0.5$ is the $0.82$-th quantile;
for the cases where people indicated \emph{presence} of respiratory diseases $p=0.5$
is the $0.675$-th quantile. 
On the broader population, this bias should be even more salient, but we must take into account that
the collected data might have more COVID-positive cases or other acute respiratory diseases
than average due to the interest in the application in the first days of the release.
From this observation, we may expect that our model's probability distribution $p(x)$ is
not so far from the true distribution $q(x)$ of having COVID.
Whereas it is unclear how to estimate $q(x)$ better than the Bernoulli distribution of either having COVID or not, it is clear that $q(0) \gg q(1)$, i.e., the probability that a random person in the population is not affected by COVID is much greater than the opposite.
Since our models trained on balanced input sources, frequencies, labels, etc., we suggest that the aforementioned bias towards lower probabilities validates that our model captures real distribution.
On the other hand results of our models on the crowdsourced data are on par with those
on the EPFL dataset. In section \ref{sub:perfromance_open_data} we concluded that for
the mentioned dataset such a performance is a consequence of weak labels. 
We would like to suggest that the same issue might be the reason for the poor
performance observed on our crowdsourced data.
%
%


It is important to mention that we were not able to restrict the health status self-reporting specifically to ``COVID'' due to
the application store rules.


\section{Application}
\label{sec:application}

\subsection{Cough detection and Segmentation}  
\label{sub:cough_detection}

Collected cough datasets and audio data coming from users of a diagnostic application have
a number of issues that could prevent a correct diagnosis. Unlike cough sounds collected in
controlled environments, crowdsourced audio samples and live data collected from users of
the application may be severely contaminated with background sounds, e.g. music, speech,
environmental noise, or feature audio of entirely irrelevant audio events, such as laughter,
clapping or snoring.
Besides contamination, data from different sources and datasets may have a varying audio quality
due to subsampling, frequency filtering, and~/~or lossy compression, all of which adversely
affect its spectral properties.
At the same time, there is high variability in cough sounds even among the relevant ``clean''
samples themselves. Apart from isolated wheezes, throat clears, or strong exhalations,
recordings might feature light coughs, and in many cases heavily clipped samples due to
close proximity to the microphone, or partially cut-off cough events at the start or end
of the recording, e.g. the burst without a subsequent noisy airflow.

It is, therefore, necessary to require a certain quality of the recorded respiratory event
prior to feeding it into the COVID-19 diagnosis model, \citep{imran_ai4covid-19_2020}.
In order to make a correct diagnosis possible, while not overly inconveniencing the user,
the system for determining the quality of a recording and detecting the presence of a cough
event should strike a balance between false positive and false negative rates. A particular
implementation may prompt the user for another attempt in the instance of a rejected recording, alongside a display of general
instructions outlining the recommended distance to the device or the level of background
noise (sec.~\ref{sub:implementation_details}).
%

Since the recordings of coughs are the most important input to our model, we train and fine-tune
a \emph{MobileNetV2} network, \citep{sandler_mobilenetv2_2018}, for a cough detection task
on manually labeled cough and non-cough data (see appendix~\ref{app:method_cough_validator}
for training details).
The architecture was chosen for its inference speed, size, and arithmetic complexity suitable
for deployment on mobile devices.
The dataset was compiled from public and proprietary sources (Table~\ref{tab:validator_data}).
Open datasets included Coswara~\citep{sharma_coswara_2020}, Covid19-Cough~\citep{project_fkthecovid_dataset_2021},
COUGHVID~\citep{orlandic_coughvid_2020}, Virufy~\citep{virufy}, and FSD50K~\citep{fonseca_fsd50k_2020},
while proprietary data was collected from call-centers, patient recordings made by hospital
staff, and our mobile application.
%
\begin{table*}[t!]
\caption{
    Manually labeled data used to train \emph{MobileNetV2} cough detection model.
    Open-source datasets are shown in blue; our own collected data are shown in orange.
}
\label{tab:validator_data}
\centering
\resizebox{\textwidth}{!}{%
\begin{tabular}{@{}lrrrrrrr@{}}
\toprule
    Sources & Accept & Reject & Light Cough & Clipping & Throat Clear & Inaudible & \textbf{Total} \\
\midrule
    \rowcolor[HTML]{D7EFFF} 
        Coswara               & 225   & 57    & 5   & 2   & 1    & 2   & \textbf{292} \\
    \rowcolor[HTML]{D7EFFF} 
        Covid19-Cough         & 991   & 173   & 31  & 68  & 13   & 7   & \textbf{1 283} \\
    \rowcolor[HTML]{D7EFFF} 
        CoughVid              & 1 633 & 1 487 & 56  & 52  & 39   & 15  & \textbf{3 282} \\
    \rowcolor[HTML]{D7EFFF} 
        Virufy                & 47    & 0     & 0   & 0   & 0    & 0   & \textbf{47} \\
    \rowcolor[HTML]{D7EFFF} 
        FSD labeled ``cough'' & 539   & 171   & 1   & 1   & 101  & 0   & \textbf{813} \\
    \rowcolor[HTML]{D7EFFF} 
        FSD miscellaneous     & 0     & 269   & 0   & 0   & 0    & 0   & \textbf{269} \\
    \rowcolor[HTML]{FFE1C4} 
        Call-center           & 294   & 2 298 & 2   & 1   & 5    & 0   & \textbf{2 600} \\
    \rowcolor[HTML]{FFE1C4} 
        Hospital collection   & 601   & 38    & 10  & 2   & 11   & 0   & \textbf{662} \\
    \rowcolor[HTML]{FFE1C4} 
        App data              & 744   & 209   & 10  & 34  & 14   & 4   & \textbf{1 015} \\
\midrule
    \textbf{Total} & \textbf{5 074} & \textbf{4 702} & \textbf{115} & \textbf{160} & \textbf{184} & \textbf{28} & \textbf{10 263} \\
\bottomrule
\end{tabular}%
}
\end{table*}

\begin{figure}[!t]
    \centering
    \includegraphics[width=0.45\textwidth]{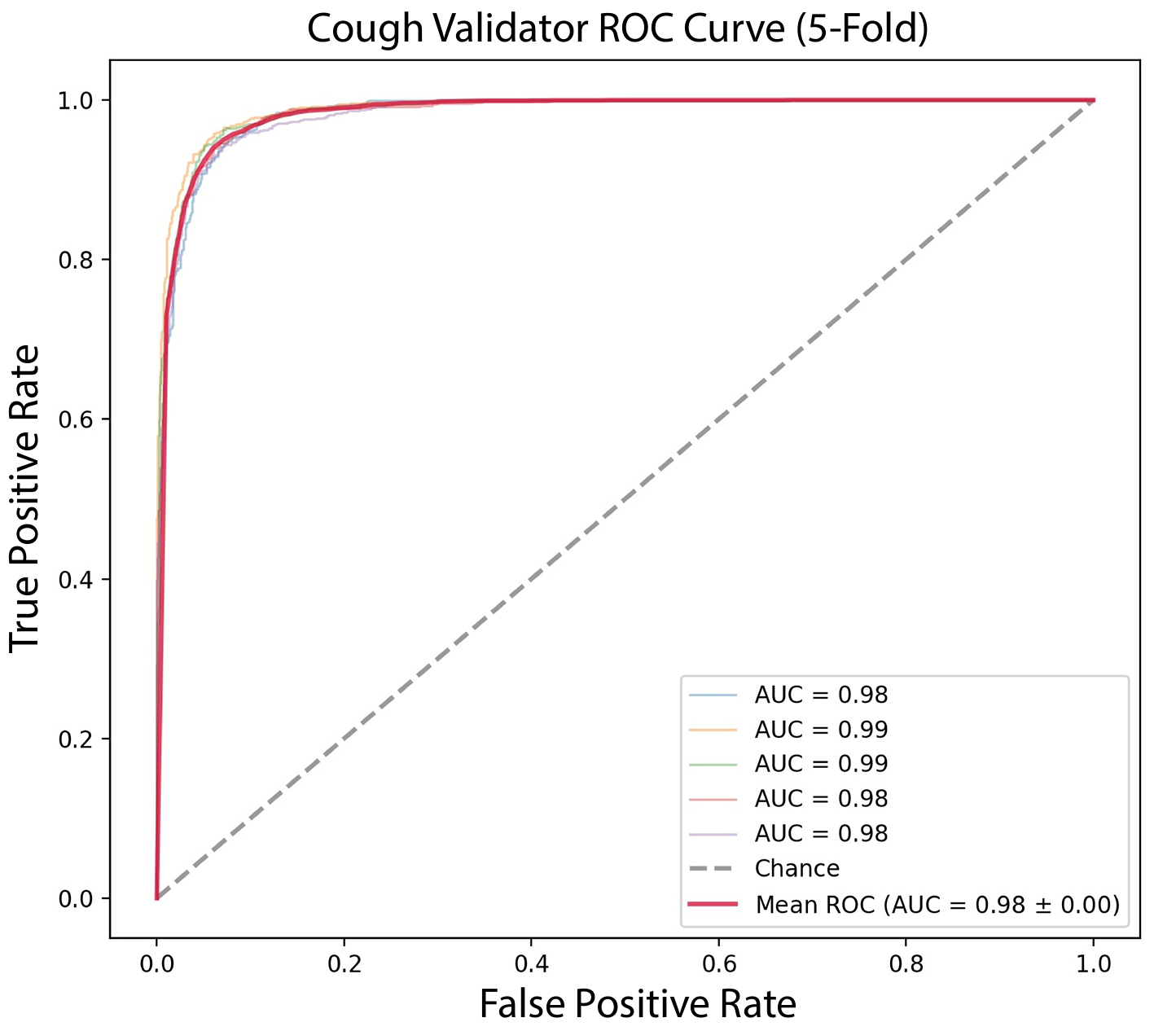}
    \caption{
        Receiver operating characteristic curve of the cough validator.
    }
    \label{fig:ROC_validator}
\end{figure}

Preprocessing was done in python with the \emph{librosa} package to trim and extract continuous
non-silent audio intervals, that were then manually segmented. Ground truth event labeling
was done by a single party to ensure consistency -- subjective thresholds were set for the
lightness of coughs and the amount of audio clipping distortion before a sample was marked
as rejected. Call-center recordings included the full conversation and therefore yielded many
rejected segments.
%
%
A total of $10 263$ recordings were labeled, of which $5 074$ were cough sounds; the remaining
$5 189$ non-cough samples consisted of non-cough events as well as the undesirable cough
variations described above.
Figure~\ref{fig:ROC_validator} depicts the ROC curves resulting from $5$-fold cross-validation
of the cough detection model trained in the pooled dataset (Table~\ref{tab:validator_data}).
Segmented recordings were labeled only with a cough or non-cough flag; disjoint user sets
across folds were not enforced.

\subsection{Performance on real data}
\label{sub:performance_on_real_data}
\begin{figure}[!t]
    \centering
    \includegraphics[width=0.48\textwidth]{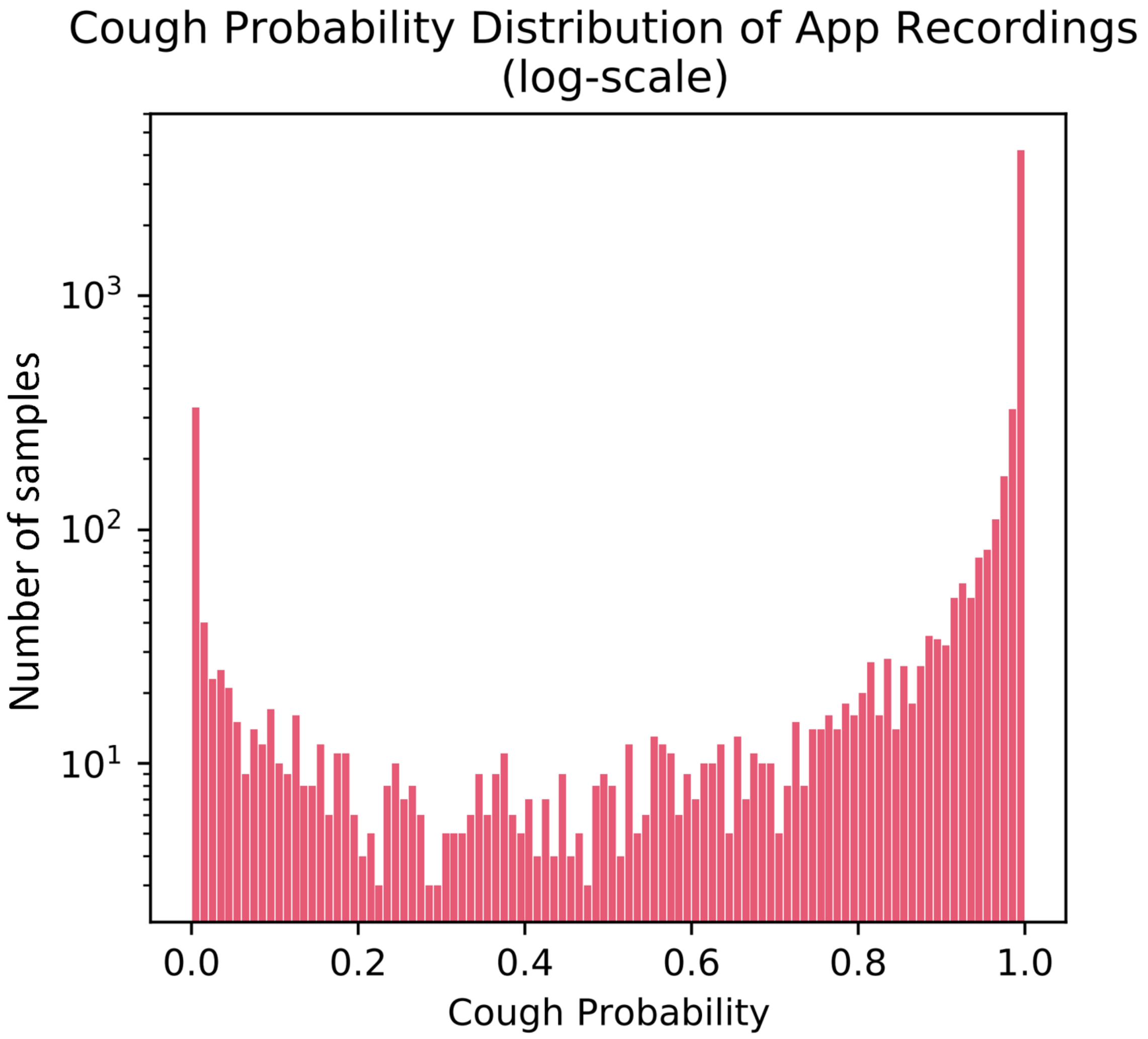}
    \caption{
    Distribution of cough validator model outputs for 6480 recordings collected by the mobile application over several days following release.}
    \label{fig:histogram_validator}
\end{figure}

We evaluated the behavior of the cough detection model on the data collected by the mobile application. The threshold for the model was set at \emph{reject $< 0.25 \leq$ accept} to give some leeway to users who might record unusual coughs, while still filtering out clearly undesirable samples and prompting for a cleaner recording. The distribution of model outputs representing the probability of a cough in a given recording is shown in Figure \ref{fig:histogram_validator}, with $9.8\%$ of recordings being rejected, i.e. classified as not containing a cough. 

We also considered the user response to receiving a prompt to re-record a sample. A re-recording of rejection was defined as an initially rejected sample followed up by another recording by the same user within 20 minutes of the first. Out of the rejected recordings, $37\%$ were re-recorded.

To examine whether the recording instructions shown by the mobile application (alongside a request to re-record) were helpful in improving the quality of the samples, we evaluated the number of successful attempts by users to produce an accepted sample, in terms of re-recording sequences. A rerecording sequence begins with a rejected recording and consists of two or more recordings that follow within 20 minutes of each other. The sequence ends when an accepted recording is achieved, or when the user does not make another recording within 20 minutes of the last. A successful sequence ends with an accepted recording, while an unsuccessful sequence ends with a rejected recording. Out of all re-recording sequences, $68\%$ were successful, showing that in the majority of instances where users attempted to correct an unsatisfactory recording, they were able to do so.

We believe that implementing similar filtering mechanisms with an opportunity to re-submit a sample would be beneficial for all audio collection efforts of this sort, especially crowdsourcing campaigns, as it would promote standardization of samples and produce larger and cleaner datasets. Users or volunteers would generally be inclined to receive an accepted status on their submitted data, as long as the re-recording process is quick and simple. By retaining the rejected samples, such data collection can only produce more samples than without filtering, while still requiring the same level of commitment from each user at the outset.

\begin{table}[!t]
    \caption{ROC AUC and Matthews Correlation Score for different ensembles on data collected from the app. Answers to the question "Do you have acute respiratory disease right now?" were considered as ground truth.}
    \label{tab:performance_on_app_data}
    \centering
    \begin{tabular}{l|r|r}
    \toprule
        Model                    & ROC AUC         & MCC            \\
    \midrule
        CNN                      & 0.6167          & 0.132          \\
        GB                       & 0.6162          & \textbf{0.148} \\
    \midrule
        GB-breath                & 0.502           & 0.021          \\
        GB-vocalization          & 0.482           & 0.017          \\
    \midrule
        Ensemble (Variant   I)   & 0.5547          & 0.111          \\
        Ensemble (Variant  II)   & \textbf{0.6227} & 0.147          \\
    \bottomrule
    \end{tabular}
\end{table}

\subsection{Application Implementation}  
\label{sub:implementation_details}

The application is implemented in client-server manner with iOS~/~Android UI frontend
and server-side storage and computational backend, implemented in \emph{Flask}.\footnote{
    Web application framework, \url{https://flask.palletsprojects.com/}
} The trained deep convolutional networks are converted into \emph{ONNX format}\footnote{
    Open Neural Network Exchange, \url{https://onnx.ai/}
} and operate in inference mode using the \emph{ONNX Runtime}\footnote{
    High-performance inference engine for ML, \url{https://www.onnxruntime.ai/}
} library for optimal throughput.
%
During each user session, the client app collects the data (symptoms, samples of cough,
breath and voice) and sends it to the server, which processes it, applies the models
and yields the response. The backend instances are replicated between several servers
with a load balancer distributing requests and workload among them for fault tolerance.

The user flow through the application requires them to record breath and cough samples
but allows them to opt out of the voice recording step. The client also prompts the user
for another cough recording if the detector is unable to spot any cough-events in
the submitted audio sample.
%

\section{Broader impact and mass testing considerations}
\label{sec:considerations}

A data-driven mass testing tool can be an invaluable and low-cost solution to identifying
disease carriers in the population and encouraging these individuals to self-isolate
\citep{larremore_test_2021}. It can provide real-time data on infection hotspots and
inform the allocation of healthcare resources.
At the same time, the question of trust in machine learning models and algorithms that
have policy implications or critically affect personal decision making is very pertinent
in medical applications, especially since most AI tools are based on statistical analysis,
rather than causation. Due care must be taken to clearly communicate and ensure the users'
awareness that the outcome of an uncertified ML-based solution \emph{does not constitute
medical advice}.

Low sensitivity or specificity in such tools can exacerbate the spread of disease,
\citep{gray2020no}. For example, a high false positive rate erodes trust in pre-screening compelling the users to brush off the alerts, which in the case of
a true positive alert fails to stimulate urgency to self-isolate or seek medical advice.
Excess trust in a tool with a high false negative rate carries the danger of conveying
a false sense of security to those who are shown a negative result, \citep{Wisem4690}.
In this case, COVID carriers, who receive such a diagnosis, might choose to forego clinical
screening methods and continue their social interactions, even when experiencing mild
symptoms. Others might neglect precautionary protective measures if they are confident
in the negative test results of their social circle, \citep{coppock_end--end_2021}.
Conversely, an unduly trusted model with a high false positive rate can overwhelm
the healthcare system, or cause overreaction in the form of severe epidemic control
measures that drastically impact individuals and businesses, \citep{surkova2020false}.
%

To mitigate the adverse societal effects of misplaced trust, we take care to communicate
to the user of the application that the application is not a certified medical tool, and
encourage them to exercise caution and seek proper medical advice or clinical testing,
such as an RT-PCR test.

\section{Limitations}
\label{sec:limitations}

The development of a model that predicts whether a recorded cough has signs of a respiratory
disease imposes strict restrictions on how data should be collected, especially if one is
working with crowdsourced data. For instance, the results of Table \ref{tab:performance_on_app_data}
are of a limited interpretation due to the absence of the ground truth labels for crowdsourced
records. For a fair comparison between models, one should collect massive crowdsourced data
verified by PCR test. This would crucially enable the identification of asymptomatic individuals
who would otherwise not go through clinical testing. 

We addressed some possible biases caused by different data sources by dividing the dataset into groups with the same properties, such as sample rate and device type. During training, objects were sampled from each group in such a way that the weights of the positive and negative classes within each group were equal. It is important to note, however, that the models were fine-tuned with cough samples that could introduce other kinds of biases. Positive COVID recordings were collected from patients in hospitals, which
constitute a relatively noisy and echo-prone environment. COVID-free recordings were sampled
from an office location, which can be expected to be generally quieter. Moreover, the participants
in the office setting might have been compelled to cough more lightly than hospital patients
and were likely on average to be younger than those admitted to medical care. These distinct
conditions and confounders create the potential for bias in the models that exploit acoustic
characteristics of the samples unrelated to features of COVID coughs. 


Our model is also limited only to the detection of coughs characteristic of COVID; we must extend our
data in order to detect other pathologies.

\section{Conclusion}
\label{sec:conclusion}
%

Our application is an attempt to make the identification of people affected by COVID easier and faster. 
We received a lot of help from the medical society at large to develop our app. 
This help came from doctors who helped us collect data, as well as from heads of clinics
and other management who suggested we use their data and collaborate in order to collect
high-quality samples. 
This shows the widespread necessity of such a service.

We encountered several projects similar to ours.
Some of them were concentrated on collecting data and sharing it with the scientific community. 
Often these datasets were very noisy and models trained on them had poor generalisation capabilities.
Some works focused on a method that could maximize the performance of a model on private data.
Our application serves both these tasks: it is able to collect data and make a prediction.

In this work, we contribute to the scientific community by providing baselines on open datasets,
and describing our method that combines feature engineering, classical and deep machine learning
methods. 
Another result of our work is the mobile app. 

The further work is twofold.
First, we will continue to collect data from healthy people and people affected by COVID.
We hope that models trained on new massive and diverse data will be more robust.
Obviously, COVID is not the only respiratory disease that might be detected.
The detection of new diseases is possible in the presence of corresponding datasets.
This is the second direction of further work: collecting data corresponding to different
respiratory diseases and training or fine-tuning models on this new data.
Based on our model we released a mobile application for public use \citep{sbermedai_app_2021},
which is available on the
App Store  
and Google Play.  

\appendices

\section{Cough detection model}
\label{app:method_cough_validator}
The cough detection model was trained using the PyTorch library and a Tesla K80 GPU.
The \emph{MobileNetV2} model was modified to accept a single-channel input for greyscale
Mel-spectrograms. The segmented cough recordings were first normalized by peak absolute
value, then downsampled to $8$~kHz, and padded to $2$ seconds if necessary. Mel-scaled
spectrograms were produced using the \emph{librosa} package with FFT window length of
$743$ samples, hop length of $186$, and the number of Mel-frequency bins of $128$.
Bootstrapping was used during training, with random crops to generate $128\times512$
spectrograms. The model weights were randomly initialized and the data was randomly
split into $80\%$ training and $20\%$ validation sets. The training was done with Adam
optimizer, an initial learning rate of $10^{-3}$, and a cosine annealing scheduler
($10$ max iterations, $5 \cdot 10^{-6}$ min learning rate). The batch size was set
at $8$ samples, and the model was trained with early stopping on the validation split
and $p=0.2$ dropout. The output layer of the model used sigmoid activation and the loss
criterion used was Binary Cross Entropy.

\section{Failed Approaches} 
\label{app:failed_models}
We were not able to obtain any improvement using Poisson masking from \citep{laguarta_covid-19_2020}.
Similar to Poisson masking we tried to utilize gradient masking reducing the importance of
high frequencies, but the attempts were unsuccessful. We were unable to find an effective
augmentation scheme for the training of the ensemble of deep neural networks.



%
%
We did not seek certification for our application.


    
    

\ifCLASSOPTIONcaptionsoff
  \newpage
\fi

\bibliography{references}  


\end{document}